\documentclass{article}

\usepackage{microtype}
\usepackage{graphicx}
\usepackage{subcaption}
\usepackage{booktabs} 

\usepackage{hyperref}

\usepackage[preprint]{icml2026}

\usepackage{amsmath}
\usepackage{amssymb}
\usepackage{mathtools}
\usepackage{amsthm}

\usepackage{booktabs}
\usepackage{ulem}
\usepackage{multirow}
\usepackage{makecell}

\usepackage[capitalize,noabbrev]{cleveref}

\theoremstyle{plain}

\theoremstyle{definition}

\theoremstyle{remark}

\usepackage[textsize=tiny]{todonotes}

\icmltitlerunning{Few-Shot Distribution-Aligned Flow Matching for Data Synthesis in Medical Image Segmentation}

\begin{document}

\twocolumn[
    \icmltitle{Few-Shot Distribution-Aligned Flow Matching for Data Synthesis in \\Medical Image Segmentation}



    \icmlsetsymbol{corres}{*}

    \begin{icmlauthorlist}
        \icmlauthor{Jie Yang}{whu,sii}
        \icmlauthor{Ziqi Ye}{sii}
        \icmlauthor{Aihua Ke}{whu}
        \icmlauthor{Jian Luo}{whu}
        \icmlauthor{Bo Cai}{whu,corres}
        \icmlauthor{Xiaosong Wang}{shailab,sii,corres}
    \end{icmlauthorlist}

    \icmlaffiliation{whu}{Key Laboratory of Aerospace Information Security and Trusted Computing, Ministry of Education, School of Cyber Science and Engineering, Wuhan University, Wuhan, China}
    \icmlaffiliation{sii}{Shanghai Innovation Institute, Shanghai, China}
    \icmlaffiliation{shailab}{Shanghai AI Laboratory, Shanghai, China}

    \icmlcorrespondingauthor{Bo Cai}{caib@whu.edu.cn}
    \icmlcorrespondingauthor{Xiaosong Wang}{xiaosong.wang@live.com}

    \icmlkeywords{Generative Model, ICML}

    \vskip 0.3in
]



\printAffiliationsAndNotice{}  

\begin{abstract}
    Data heterogeneity hinders clinical deployment of medical image analysis models, and generative data augmentation helps mitigate this issue. However, recent diffusion-based methods that synthesize image-mask pairs often ignore distribution shifts between generated and real images across scenarios, and such mismatches can markedly degrade downstream performance. To address this issue, we propose AlignFlow, a flow matching model that aligns with the target reference image distribution via differentiable reward fine-tuning, and remains effective even when only a small number of reference images are provided. Specifically, we divide the training of the flow matching model into two stages: in the first stage, the model fits the training data to generate plausible images; Then, we introduce a distribution alignment mechanism and employ differentiable reward to steer the generated images toward the distribution of the given samples from the target domain. In addition, to enhance the diversity of generated masks, we also design a flow matching–based mask generation to complement the diversity in regions of interest. Extensive experiments demonstrate the effectiveness of our approach, i.e., performance improvement by  3.5-4.0\% in mDice and 3.5-5.6\% in mIoU across a variety of datasets and scenarios.
\end{abstract}

\section{Introduction}
\label{sec:intro}

\begin{figure}
    \centering
    \centerline{\scalebox{0.52}{\includegraphics[width=\textwidth]{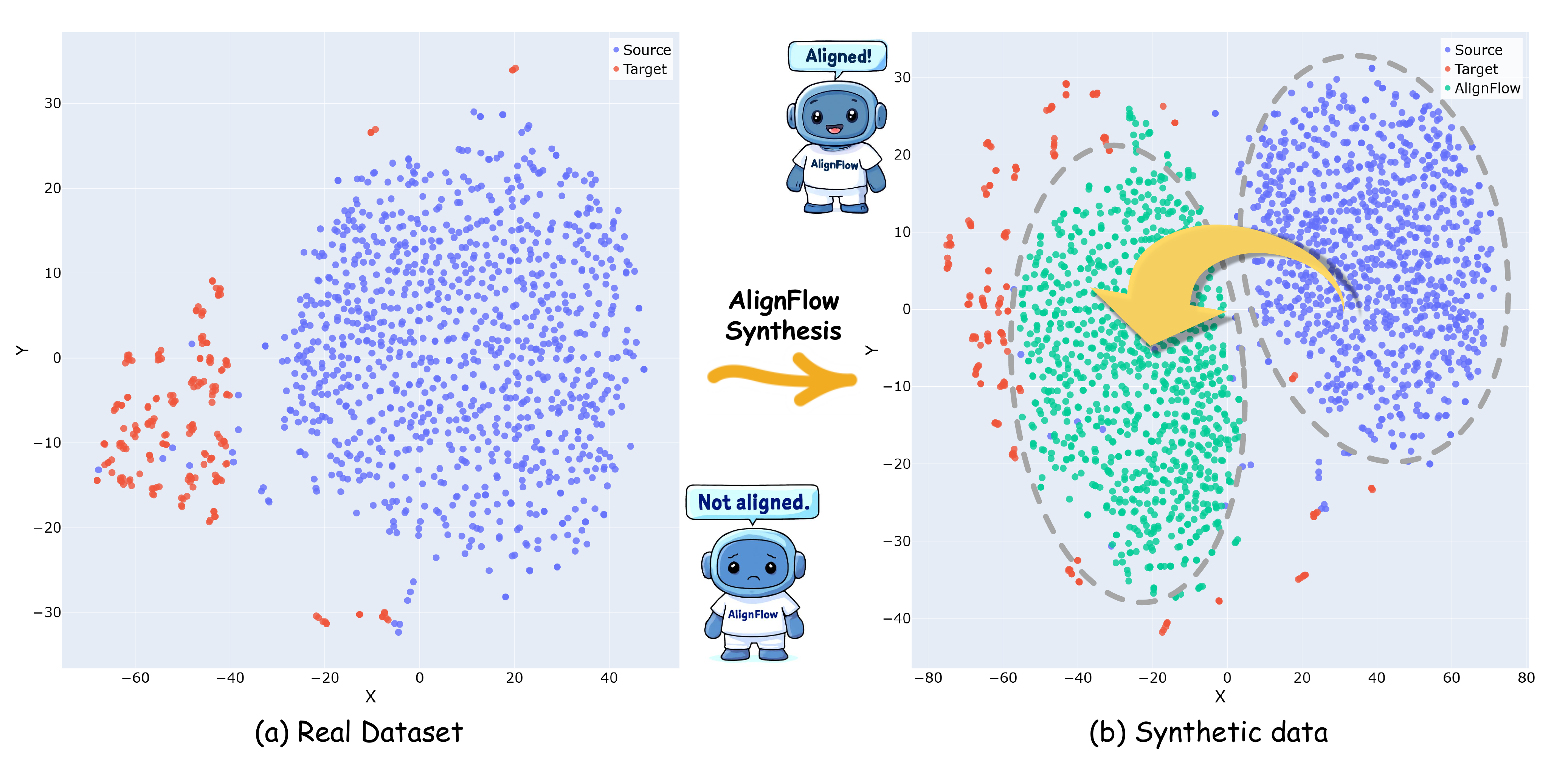}}}
    \caption{Illustration of the data distribution of images generated by AlignFlow that align with the target domain distribution. The green dots represent the data points of images generated by AlignFlow, which are accurately scattered at the center of the target-domain data points.}
    \label{fig:fig1}
\end{figure}

With the rapid advancement of artificial intelligence (AI) models over the past decade, medical image segmentation has achieved remarkable progress. However, constrained by limited data amount and diversity, medical image segmentation models have not yet reached their full potential compared to their counterparts for natural images. To overcome the limitation of insufficient training data, various solutions have been proposed, ranging from early traditional data augmentation methods \cite{zhang2017mixup, zhong2020random, yun2019cutmix, ghiasi2021simple}, to subsequent GAN-based image synthesis approaches \cite{nie2017medical, costa2017end, qi2020sag}, and more recently, to the mainstream diffusion-based image synthesis methods \cite{xu2024medsyn, wang2025self, wang20253d}. Although these methods have effectively enhanced the raw quality of generated images, real-world clinical applications may require further specification and customization, e.g., region of interest (ROI) and domain shift.

In synthesizing medical image segmentation datasets, it is particularly essential that the generated images accurately correspond to the masks, meaning that the mask's size, shape, and location must align with the lesions or organs in the generated image. This requires incorporating additional mask control information during image generation. The ControlNet \cite{zhang2023adding} architecture is a widely used framework for prompt-controlled image generation. It employs an external network to encode the mask and injects the mask control information into a frozen diffusion model via residual connections, thereby enabling the generation of images with ROIs that are structurally consistent with the mask. To further balance both the fidelity and diversity of generated images, Siamese-Diffusion \cite{qiu2025noise} introduces a dual-branch model comprising an image diffusion branch and a mask diffusion branch. During training, a noise consistency loss is used to allow the mask diffusion branch to learn from the image diffusion branch, enhancing the fidelity of mask-based generation.

Despite these advancements, few studies have considered the more detailed discrepancy between the data distribution of generated images and that of images acquired in real-world scenarios. As shown in Figure \ref{fig:fig1}(a), medical imaging devices of different brands, models, and parameter settings exhibit distinct imaging characteristics. Moreover, medical images vary considerably due to differences in patient age, gender, and ethnicity. Consequently, when a model trained on data from medical institution A is applied to institution B, its performance often declines to some extent. This decline arises from the distribution mismatch between the training data and the real-world data (hereafter referred to as target domain data).

To tackle this issue, we hypothesize that generating images that are more consistent with the target domain data distribution could reduce the gap between the training and target domain set. Since acquiring target domain data is often challenging, it would be ideal if this goal could be achieved using only a small number of target domain samples. Motivated by this, we propose AlignFlow, a flow-matching model that employs differentiable reward to enforce the distribution alignment of generated images with reference images, and remains effective even when only a few reference images are provided.

During the training of AlignFlow, we divide the process into two stages. In the first stage, we optimize only the denoising loss, enabling the model to generate images consistent with the training dataset distribution. In the second stage, we jointly optimize the denoising loss and the alignment loss, allowing the model to retain its original generative capability while aligning the generated images with the reference images. To accurately compute the alignment loss, we introduce the following technical components: 1) We design a differentiable reward function and optimize model parameters directly via differentiable reward fine-tuning using gradient updates. 2) We develop a partially tunable inference strategy, which allows the model to store only a single-step inference computational graph during training, thereby avoiding memory overflow and gradient explosion issues associated with multi-step inference. 3) We use DINOv3 to extract global pooled image features, leveraging its strong prior knowledge to make precise judgments about image distributions. 4) We propose a reward function based on Maximum Mean Discrepancy (MMD)\cite{gretton2012kernel} to measure the discrepancy between two image distributions effectively. Additionally, to further enhance the diversity of masks, we propose a mask synthesis method that includes a series of post-processing and filtering steps, enabling pre-trained flow matching models—originally designed for natural image generation—to produce high-quality masks.

The main contributions of this paper is three-fold:
\begin{itemize}
    \item We propose AlignFlow, a few-shot distribution-aligned flow matching model that uses a small set of reference images to generate synthetic data matching the target-domain distribution. Combined with our mask synthesis method, it further increases data diversity.
    \item We propose a distribution alignment mechanism and design two reward functions that efficiently assess the differences between generated and reference images.
    \item We conducted extensive experiments on six datasets from different domains and three segmentation models, demonstrating that AlignFlow significantly outperforms all previous generative methods.
\end{itemize}

\section{Related Work}
\label{sec:relatedwork}

\begin{figure*}[!t]
    \centering
    \centerline{\includegraphics[width=\textwidth]{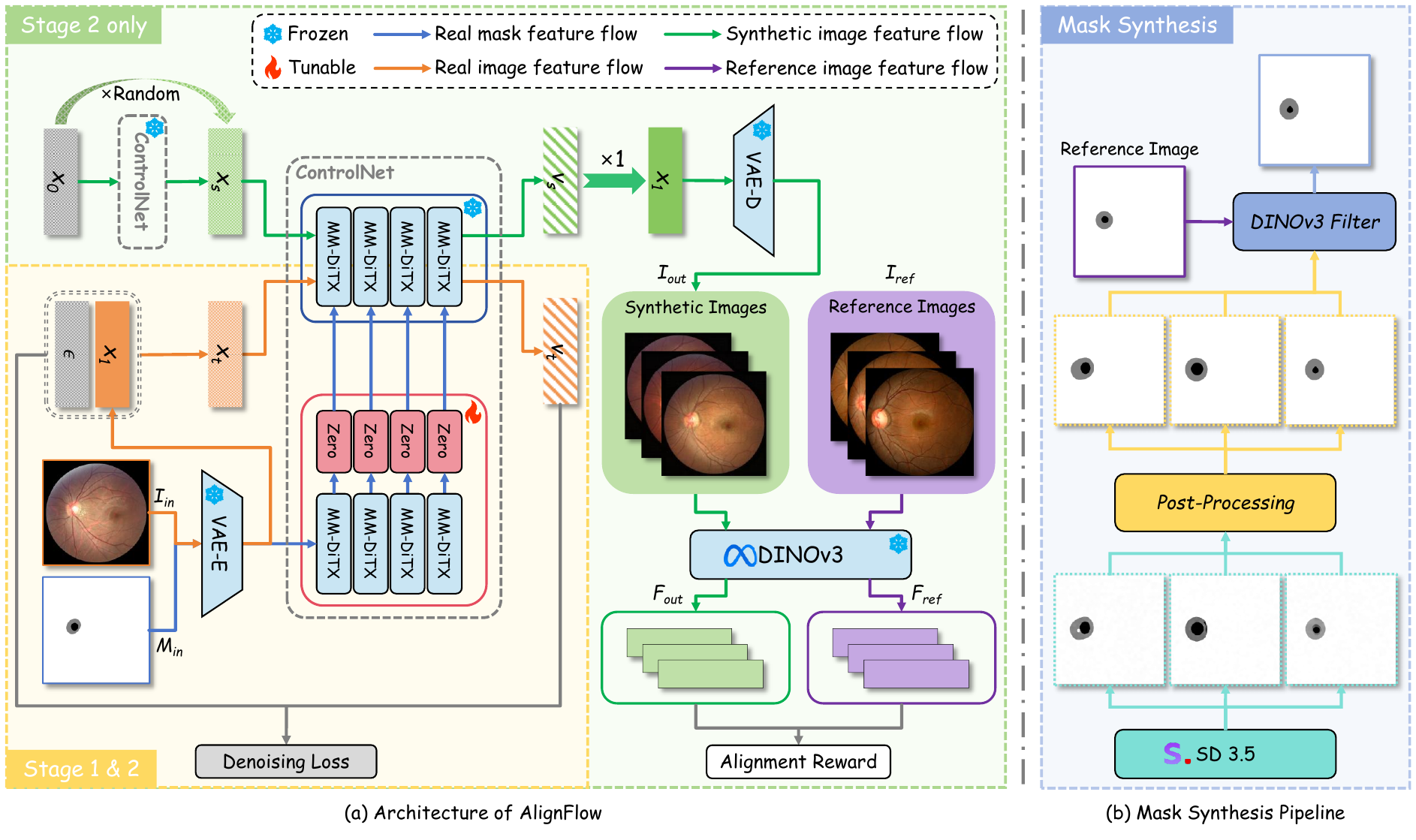}}
    \caption{(a) Illustration of the AlignFlow architecture. In stage 1, we optimize the denoising loss to enable the model to generate reasonable images; in stage 2, we simultaneously optimize the denoising loss and the alignment loss, allowing the model to align the generated images with reference images from the target domain while maintaining its original image generation capability. (b) Illustration of the mask synthesis pipeline.}
    \label{fig:fig2}
\end{figure*}

\noindent\textbf{Flow Matching.} Although diffusion models \cite{ho2020denoising, song2020denoising, song2020score, rombach2022high} have achieved significant progress in image generation, they still suffer from long training times and low sampling efficiency. The introduction of flow matching \cite{lipman2022flow, liu2022flow} effectively addresses these limitations. Flow matching is a training method based on Continuous Normalizing Flows, which extends beyond the traditional diffusion model framework by allowing the use of a broader range of probability paths, including optimal transport. Due to the characteristics of optimal transport probability paths—such as linear simplicity and a constant vector field direction—flow matching is more efficient in training than conventional diffusion probability paths, while also producing higher quality samples. Subsequent works \cite{esser2024scaling, domingo2024adjoint, bandyopadhyay2025sd3} have applied modern techniques popular in diffusion models to flow matching, gradually establishing it as the de facto standard in high quality image generation.

\noindent\textbf{Diffusion Post-Training.} In natural image diffusion generative models, reinforcement learning \cite{clark2023directly, prabhudesai2023aligning, prabhudesai2024video, xu2023imagereward} serves as a mainstream post training method and is widely used for human preference alignment and for improving the numerical accuracy of generated content. DDPO \cite{black2023training} models the diffusion denoising process as a multi step Markov decision process and applies policy gradients to directly optimize external rewards along generation trajectories for goal directed fine tuning. Flow-GRPO \cite{liu2025flow} introduces GRPO based online optimization into flow matching model training, converts deterministic sampling into a stochastic process, and reduces training steps to improve efficiency and alignment performance. However, because existing diffusion generative models are pretrained on large scale natural image datasets, they exhibit sufficient generalization to produce outputs that are likely to elicit positive feedback from reward models, so these methods demonstrate strong performance in the natural image domain, but they are almost unusable in medical image generation settings with limited data.

\section{Method}
\label{sec:method}

AlignFlow uses a pre-trained flow matching-based Stable Diffusion 3.5 (SD 3.5)\cite{esser2024scaling} as the foundation model and employs ControlNet \cite{zhang2023adding} to incorporate mask control information. The training process of AlignFlow can be divided into two stages. In the first stage, we target only the denoising loss, enabling the model to generate images consistent with the training data. In the second stage, we compute both the denoising loss and the alignment loss, allowing the model to align the distribution of generated images with that of the given reference images while preserving its original generative capability.

During the training of AlignFlow, we adopt a training strategy similar to that of ControlNet, namely freezing the parameters of the VAE and MM-DiTX while only fine-tuning the parameters of the ControlNet module. This allows the model to retain as much as possible the generative capabilities with good generalization acquired during large scale pre-training. For prompt selection, we use the template "An image of ..." to encode image descriptions. As shown in Figure \ref{fig:fig2}(a), in the first stage of training, we employ only the standard flow matching training strategy. Given an input image $I_{in}$ and a mask $M_{in}$, they are first encoded by the VAE encoder $\varepsilon$ into corresponding latent features $x_1$ and $c_m$, while the prompt is encoded by a text encoder into the latent feature $c_p$. The mask latent $c_m$ and text latent $c_p$ are then input into the model, where features are first extracted using the MM-DiTX copied from SD 3.5, and then gradually injected into the frozen SD 3.5 through zero-initialized linear layers. The image latent $x_1$ is linearly interpolated with random Gaussian noise $\epsilon$ to obtain the noisy latent $x_t$. After passing through SD 3.5, $x_t$ is used to predict the vector field $v_t$ at time $t$, and the denoising loss $L_D$ is computed according to the following formula:
\begin{align}\label{eq:eq5}
     & L_D = \mathbb{E}_{t,x_1,\epsilon \sim N(0,1)}[\Vert v^\theta_t - (x_1-\epsilon) \Vert^2] \\
     & \ v^\theta_t = v^\theta_t(x_t,t,c_m,c_p), \ x_t = tx_1 + (1-t)\epsilon
\end{align}
where $\theta$ denotes the parameters of ControlNet. In the second stage of training, in addition to the denoising loss, we introduce a distribution alignment mechanism. A differentiable alignment reward is designed to score the distribution discrepancy between the generated images and the given reference images. By taking the negative of this reward function, we obtain the alignment loss $L_A$. The alignment loss $L_A$ is then weighted by $w_{align}$ and combined with the denoising loss $L_D$ for the second-stage training. The loss function $L$ is formally expressed as follows:
\begin{align}\label{eq:eq6}
    L = \left\{
    \begin{aligned}
         & L_D, \                 & stage \ 1 \\
         & L_D + w_{align}L_A, \  & stage \ 2
    \end{aligned}
    \right.
\end{align}

\subsection{Distribution Alignment}
As mentioned above, the distribution gap between training data and real-world data can reduce the model’s performance on real-world data. To address this issue, we propose a distribution alignment mechanism. Specifically, in the second stage of training, in addition to the standard denoising loss, we introduce a partially tunable inference process. As shown in Figure \ref{fig:fig2}(a), we first randomly sample a Gaussian noise $x_0$, and then randomly sample a time step $s$ between the start time step $t_s$ and $1$. The initial noise $x_0$ is then inferred to $x_s$ using the fully frozen ControlNet $v^{\theta^*}$ with a step size $d_t$. In our implementation, we set $d_t = \frac{1}{20}$, and the iterative process is formally expressed as follows:
\begin{align}\label{eq:eq7}
    x_{t+d_t} & = x_t + d_t \cdot v^{\theta^*}_t(x_t,t,c_m,c_p)
\end{align}
After inferring $x_s$, we unfreeze the previously frozen parameters to obtain $v^\theta$, and then use $v^\theta$ to perform a single-step denoising on $x_s$ to directly obtain $\hat{x_0}$. The computation process is as follows:
\begin{align}\label{eq:eq8}
    \hat{x_0} = x_s + (1-s)v^{\theta}_t(x_s,x,c_m,c_p)
\end{align}
Thanks to the straighter probability paths of flow matching, even if multiple time steps are skipped during inference, the model can still generate reasonable images. This property forms the basis for accurately computing the alignment loss. After generating the latent $\hat{x_0}$, we first decode it into an image $I_{out}$ using the VAE decoder $\varepsilon^{-1}$. We then extract global pooled features $F_{out}$ and $F_{ref}$ from $I_{out}$ and $I_{ref}$ using DINOv3, which provides strong prior knowledge. The distribution discrepancy between the two sets of features is computed using Maximum Mean Discrepancy (MMD)\cite{gretton2012kernel} as $MMD(F_{out}, F_{ref})$. Since the MMD value is negatively correlated with the distribution similarity, we define the alignment reward as $r_{align} = -MMD(F_{out}, F_{ref})$. Because model parameters are optimized via gradient descent, we take the negative of the reward function to obtain the alignment loss $L_A$, thereby satisfying the objective of maximizing the reward function. In summary, the expression for $L_A$ is as follows:
\begin{align}\label{eq:eq9}
    L_A & = -r_{align} = MMD(F_{out},F_{ref}).
\end{align}

\begin{figure*}[!t]
    \centering
    \centerline{\includegraphics[width=\textwidth]{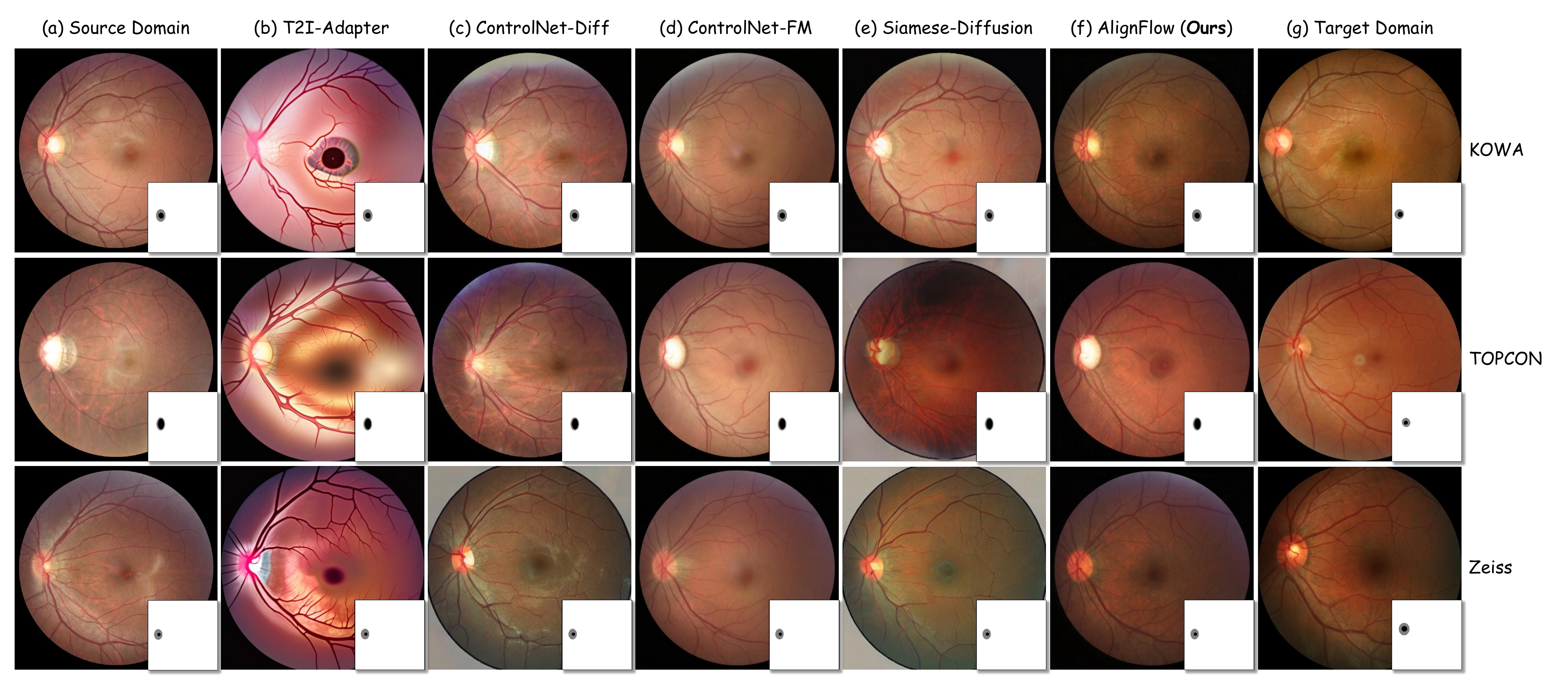}}
    \caption{Qualitative comparison on REFUGE2 dataset. The source domain is Canon, and the target domains are annotated on the right side of each row.}
    \label{fig:fig3}
\end{figure*}

\subsection{Reward Design}
To implement the reward function, we designed two schemes to investigate the impact of different approaches on AlignFlow's performance.

\noindent\textbf{Maximum Mean Discrepancy.} Maximum Mean Discrepancy (MMD) is commonly used to measure the distance between two distributions, and it is defined as follows:
\begin{align}\label{eq:eq10}
    MMD_{\mathcal{H}}(P,Q) \coloneqq \mathop{sup}\limits_{f \in \mathcal{H}} \big(\mathbb{E}_P[f(X)] - \mathbb{E}_Q[f(Y)] \big)
\end{align}
where $\mathcal{H}$ denotes a reproducing kernel Hilbert space, and $X$ and $Y$ represent random variables following distributions $P$ and $Q$, respectively. When using MMD to measure the distance between the generated image features $F_{out}$ and the reference image features $F_{ref}$, we employ the mean as an unbiased estimate of the expectation and use a polynomial kernel $f_p(x,y)$ as the specific implementation of the kernel function. Its mathematical expression is as follows:
\begin{align}\label{eq:eq11}
    f_p(x,y) = (\langle x,y \rangle / d + c)^\delta
\end{align}
where $d$ denotes the feature dimension, $c$ represents the coefficient, and $d$ indicates the degree. In our implementation, $d = 1280$, $c = 1.0$, and $d = 3$. In summary, the expression for the reward function $r_{align}^{MMD}$ is as follows:
\begin{align}\label{eq:eq12}
    \begin{aligned}
         & r_{align}^{MMD}(F_{out},F_{ref}) = -\frac{1}{M(M-1)} \sum_{i \neq j}^{M} f_p(x_i,x_j)                        \\
         & + \frac{2}{MN} \sum_{i=1}^{M}\sum_{j=1}^{N} f_p(x_i,y_j) - \frac{1}{N(N-1)} \sum_{i \neq j}^{N} f_p(y_i,y_j)
    \end{aligned}
\end{align}
where $M$ and $N$ denote the number of samples in $F_{out}$ and $F_{ref}$, respectively.

\noindent\textbf{Symmetric Kullback-Leibler Divergence.} The Symmetric Kullback-Leibler Divergence (SKL) is defined as:
\begin{align}\label{eq:eq13}
     & SKL(P \Vert Q) \coloneqq \frac{1}{2} [KL(P \Vert Q) + KL(Q \Vert P)]         \\
     & KL(P \Vert Q) \coloneqq \sum_{x \in \mathcal{X}} P(x) \log \frac{P(x)}{Q(x)}
\end{align}
In our implementation, to ensure that the number of samples in $F_{out}$ and $F_{ref}$ is consistent, we first compute the mean of $F_{out}$ and $F_{ref}$ before calculating the SKL. In summary, the expression for the reward function $r_{align}^{SKL}$ is as follows:
\begin{align}\label{eq:eq14}
    \begin{aligned}
        r_{align}^{SKL}(F_{out},F_{ref}) = & -\frac{1}{2}[KL(F_{out} \Vert F_{ref}) \\
                                           & + KL(F_{ref} \Vert F_{out})]
    \end{aligned}
\end{align}
We conducted comparative experiments across multiple datasets and segmentation models, with the results shown in Table \ref{tab:tab2}. We found that using MMD as the reward function generally achieves higher accuracy on downstream segmentation tasks compared to using SKL. Therefore, we adopt $r_{align}^{MMD}$ as the default implementation of $r_{align}$.

\subsection{Mask Synthesis}
To enhance the generalization performance of downstream segmentation models, we also propose a mask generation method to synthesize diverse masks. These new masks are then used to infer their corresponding images, which are subsequently employed to train the downstream segmentation models. Similarly, we adopt SD 3.5 as the foundation model as shown in Figure \ref{fig:fig2}(b), taking the mask as the input and fine-tuning the model via the denoising loss. However, the directly generated images often contain details inconsistent with real masks, including: 1) Pixels in the mask that do not belong to the specified class; 2) Jagged textures along mask edges, resulting in insufficient smoothness; 3) Isolated pixels of other classes appearing within objects or the background of the mask; 4) Excessive deviations between the generated mask and the real mask.

Thus, we design a series of post-processing and filtering steps: 1) Thresholding on the mask to constrain invalid pixel values within the range of the specified class. 2) Gaussian blurring to smooth the edges. 3) Opening operations to remove bright spots in the foreground and closing to remove dark spots in the background, to eliminate isolated pixels that do not belong to the current object. 4) For each real mask image, we generate $K$ synthetic mask images, extract their global pooled features using DINOv3, and compute the cosine similarity between the latent representations of the real mask and each synthetic image. For each real mask, only the one with the highest similarity is retained.

\section{Experiments}
\label{sec:experiment}

\begin{table}
    \caption{Image quality evaluation on CVC-ClinicDB dataset. "ControlNet-Diff" and "ControlNet-FM" respectively denote ControlNet models that use diffusion and flow matching as their foundation models.}
    \label{tab:tab1}
    \centering
    \setlength{\tabcolsep}{1.2mm}
    \scalebox{0.82}
    {
        \begin{tabular}{l|c c c c c}
            \toprule[2pt]
            Method                   & FID($\downarrow$) & KID($\downarrow$) & LPIPS($\downarrow$) & SSIM($\uparrow$) & PSNR($\uparrow$) \\
            \hline
            T2I-Adapter              & 226.13            & 0.2374            & 0.4426              & 0.4519           & 18.07            \\
            ControlNet-Diff          & 162.87            & 0.1378            & 0.4335              & 0.4109           & 19.61            \\
            ControlNet-FM            & 166.20            & 0.1449            & 0.4396              & 0.4394           & 20.31            \\
            Siamese-Diffusion        & 156.82            & 0.1344            & 0.4497              & 0.3982           & 19.33            \\
            \textbf{AlignFlow(Ours)} & \textbf{149.27}   & \textbf{0.1301}   & \textbf{0.3931 }    & \textbf{0.5099}  & \textbf{21.23}   \\
            \bottomrule[2pt]
        \end{tabular}}
\end{table}

\begin{table*}[t]
    \caption{Comparison of medical image segmentation performance on UNet, SegFormer, and DPT. The evaluation is performed using the mDice(\%) and mIoU(\%) metrics. "ControlNet-Diff" and "ControlNet-FM" respectively denote ControlNet models that use diffusion and flow matching as their foundation models. The "+" symbol indicates that synthetic data were added to the original training set to form the complete training dataset. The best results are highlighted in bold, while the second-best results are underlined.}
    \label{tab:tab2}
    \centering
    \setlength{\tabcolsep}{2mm}
    \scalebox{0.81}
    {
        \begin{tabular}{l|c c c c c c|c c c c c c}
            \toprule[2pt]
            \multirow{4}{*}{Methods}                           &
            \multicolumn{6}{c|}{FedPolyp\cite{chen2024fedevi}} & \multicolumn{6}{c}{REFUGE2\cite{fang2022refuge2}}                                                                                                                                                                                                                                                                              \\
            \cmidrule{2-13}
            ~                                                  & \multicolumn{2}{c|}{CVC-ClinicDB}                 & \multicolumn{2}{c|}{CVC-ColonDB} & \multicolumn{2}{c|}{ETIS} & \multicolumn{2}{c|}{KOWA} & \multicolumn{2}{c|}{TOPCON} & \multicolumn{2}{c}{Zeiss}                                                                                                                         \\
            \cmidrule{2-13}
            ~                                                  & mDice                                             & mIoU                             & mDice                     & mIoU                      & mDice                       & mIoU                      & mDice             & mIoU              & mDice             & mIoU              & mDice             & mIoU              \\
            \midrule[1pt]
            UNet\cite{ronneberger2015u}                        & 88.33                                             & 83.11                            & 79.07                     & 74.16                     & 75.46                       & 71.32                     & 89.29             & 82.14             & 84.64             & 76.96             & 87.05             & 79.23             \\
            \hline
            +T2I-Adapter\cite{mou2024t2i}                      & 88.35                                             & 83.45                            & 77.48                     & 73.11                     & 75.60                       & 72.08                     & 88.44             & 81.02             & 89.13             & 82.22             & 89.20             & 82.25             \\
            +ControlNet-Diff\cite{zhang2023adding}             & 88.06                                             & 83.25                            & 80.33                     & 75.63                     & 74.17                       & 70.67                     & 72.96             & 66.73             & 77.33             & 70.72             & 87.35             & 80.23             \\
            +ControlNet-FM                                     & 88.11                                             & 83.32                            & 80.67                     & 75.75                     & 75.75                       & 72.15                     & 88.75             & 81.30             & 87.53             & 80.28             & \textbf{90.01}    & \textbf{83.26}    \\
            +Siamese-Diffusion\cite{qiu2025noise}              & 88.11                                             & 83.38                            & 78.13                     & 73.71                     & \textbf{76.47}              & \textbf{72.91}            & 84.79             & 77.86             & 88.59             & 81.75             & 86.37             & 78.95             \\
            +AlignFlow(SKL)                                    & \textbf{88.69}                                    & \underline{83.74}                & \underline{81.29}         & \underline{76.39}         & 75.82                       & 72.09                     & \underline{89.65} & \underline{82.57} & \textbf{89.24}    & \underline{82.49} & 89.81             & 83.05             \\
            \textbf{+AlignFlow}                                & \underline{88.68}                                 & \textbf{83.77}                   & \textbf{82.41}            & \textbf{77.38}            & \underline{76.45}           & \underline{72.83}         & \textbf{89.77}    & \textbf{82.77}    & \underline{89.14} & \textbf{82.54}    & \underline{89.94} & \underline{83.22} \\

            \midrule[1pt]
            SegFormer\cite{xie2021segformer}                   & 92.15                                             & 87.92                            & 83.53                     & 79.11                     & 81.53                       & 77.83                     & 87.22             & 79.37             & 87.31             & 80.26             & 89.41             & 82.85             \\
            \hline
            +T2I-Adapter\cite{mou2024t2i}                      & 92.06                                             & 87.93                            & 82.69                     & 78.40                     & 81.45                       & 77.71                     & 88.95             & 81.70             & \underline{88.80} & \underline{82.02} & \underline{89.71} & \underline{83.02} \\
            +ControlNet-Diff\cite{zhang2023adding}             & 92.28                                             & 88.07                            & 83.97                     & 79.31                     & 81.81                       & 77.88                     & 81.12             & 73.46             & 83.50             & 76.54             & 72.37             & 65.66             \\
            +ControlNet-FM                                     & 92.65                                             & 88.55                            & 84.19                     & 79.59                     & 82.90                       & \underline{79.05}         & \underline{89.24} & 82.36             & 86.92             & 80.56             & 87.80             & 80.81             \\
            +Siamese-Diffusion\cite{qiu2025noise}              & \underline{92.77}                                 & \underline{88.64}                & \underline{85.33}         & \underline{80.38}         & \underline{82.98}           & 79.03                     & 87.11             & 80.43             & 84.62             & 77.79             & 86.86             & 79.39             \\
            +AlignFlow(SKL)                                    & 92.75                                             & 88.58                            & 84.57                     & 80.05                     & 81.70                       & 77.49                     & 89.19             & \underline{82.50} & 88.19             & 81.75             & 89.21             & 82.65             \\
            \textbf{+AlignFlow}                                & \textbf{92.87}                                    & \textbf{88.89}                   & \textbf{86.52}            & \textbf{81.34}            & \textbf{84.57}              & \textbf{80.64}            & \textbf{90.36}    & \textbf{83.96}    & \textbf{90.17}    & \textbf{83.68}    & \textbf{90.19}    & \textbf{83.85}    \\

            \midrule[1pt]
            DPT\cite{ranftl2021vision}                         & 90.03                                             & 85.13                            & 81.66                     & 76.61                     & 73.39                       & 69.76                     & 86.48             & 78.28             & 85.29             & 77.46             & 85.59             & 77.54             \\
            \hline
            +T2I-Adapter\cite{mou2024t2i}                      & 89.28                                             & 84.32                            & 82.55                     & 77.67                     & 74.39                       & 70.36                     & 87.55             & 79.67             & \textbf{85.77}    & \underline{78.19} & 85.82             & 77.87             \\
            +ControlNet-Diff\cite{zhang2023adding}             & 89.88                                             & 84.96                            & 81.80                     & 76.89                     & 71.41                       & 67.89                     & 66.75             & 59.03             & 67.98             & 60.91             & 53.28             & 48.06             \\
            +ControlNet-FM                                     & \underline{90.35}                                 & \underline{85.50}                & 82.54                     & 77.79                     & \underline{76.44}           & \underline{72.47}         & 87.67             & 79.88             & 82.96             & 75.49             & 85.53             & 77.50             \\
            +Siamese-Diffusion\cite{qiu2025noise}              & 89.11                                             & 84.30                            & 82.94                     & 78.16                     & 76.19                       & 72.11                     & 80.10             & 72.42             & 74.85             & 67.63             & 62.39             & 57.23             \\
            +AlignFlow(SKL)                                    & 90.10                                             & 85.23                            & \underline{84.59}         & \underline{79.59}         & 76.23                       & 72.21                     & \underline{87.96} & \underline{80.49} & 85.53             & 78.09             & \underline{86.47} & \underline{78.80} \\
            \textbf{+AlignFlow}                                & \textbf{90.51}                                    & \textbf{85.67}                   & \textbf{84.65}            & \textbf{79.67}            & \textbf{77.39}              & \textbf{73.22}            & \textbf{88.59}    & \textbf{81.25}    & \underline{85.70} & \textbf{78.23}    & \textbf{86.73}    & \textbf{79.09}    \\
            \bottomrule[2pt]
        \end{tabular}}
\end{table*}

\subsection{Experimental Settings}
\noindent\textbf{Datasets.} We select two types of images for experiments: gastrointestinal polyp images and retinal fundus images, each with four datasets from different domains. For gastrointestinal polyp images, following the setup of FedEvi \cite{chen2024fedevi}, we choose the Kvasir \cite{jha2019kvasir}, ETIS \cite{silva2014toward}, ColonDB \cite{tajbakhsh2015automated}, and ClinicDB \cite{bernal2015wm} datasets, which contain 1000, 196, 379, and 612 image-mask pairs, respectively. For retinal fundus images, we select the REFUGE2 dataset \cite{fang2022refuge2}, which includes images acquired by four different devices: Canon, KOWA, TOPCON, and Zeiss, with 800, 400, 400, and 400 images, respectively, each with a corresponding mask.

We use the Kvasir and Canon datasets, which contain the largest amounts of data, as source domain datasets for two different types of data, respectively, while the other three datasets corresponding to each data type serve as target domain datasets. The division of the training and testing sets follows the same scheme as described above, where we use the source domain datasets as the training set and the target domain datasets as the testing set. To ensure fairness, during training of the generative model, we first randomly sample a fixed number of images from the target domain dataset as reference images. After sampling, these images remain fixed, and we use them to guide the training process in all training stages of all compared models. In subsequent comparative experiments, we default to randomly sampling 10 reference images from each target domain dataset. For training the segmentation model, following the setup of Siamese-Diffusion\cite{qiu2025noise}, we use the masks in the training set as control masks, then perform inference with the generative model on these masks to obtain synthetic images. The training set and the generated image-mask pairs are then jointly used for training the segmentation model.

\noindent\textbf{Metrics.} We select several commonly used image quality assessment metrics to evaluate the quality of the generated images, mainly including FID\cite{heusel2017gans}, KID\cite{binkowski2018demystifying}, LPIPS\cite{zhang2018unreasonable}, SSIM\cite{wang2004image}, and PSNR. Since the goal of this study is to make the generated images match the distribution of the target domain datasets as closely as possible, we use the images from the target domain datasets as the ground truth when assessing image quality. For the segmentation model selection, we choose the commonly used CNN-based model UNet\cite{ronneberger2015u}, as well as Transformer-based models, including SegFormer\cite{xie2021segformer} and DPT\cite{ranftl2021vision}. The mDice and mIoU are used to evaluate the segmentation performance.

\begin{table}
    \caption{Validation of the effectiveness of the mask generation method, evaluated using mDice (\%) and mIoU (\%) metrics.}
    \label{tab:tab2_1}
    \centering
    \setlength{\tabcolsep}{2.0mm}
    \scalebox{0.75}
    {
        \begin{tabular}{l|c c c c c c}
            \toprule[2pt]
            \multirow{2}{*}{Methods}
            ~                & \multicolumn{2}{c|}{CVC-ClinicDB} & \multicolumn{2}{c|}{CVC-ColonDB} & \multicolumn{2}{c}{ETIS}                                                    \\
            \cmidrule{2-7}
            ~                & mDice                             & mIoU                             & mDice                    & mIoU           & mDice          & mIoU           \\

            \midrule[1pt]
            UNet             & 88.33                             & 83.11                            & 79.07                    & 74.16          & 75.46          & 71.32          \\
            \hline
            +AlignFlow       & \textbf{88.68}                    & \textbf{83.77}                   & 82.41                    & 77.38          & 76.45          & 72.83          \\
            +AlignFlow(Mask) & 87.23                             & 82.20                            & \textbf{82.52}           & \textbf{77.47} & \textbf{77.99} & \textbf{73.60} \\

            \midrule[1pt]
            DPT              & 90.03                             & 85.13                            & 81.66                    & 76.61          & 73.39          & 69.76          \\
            \hline
            +AlignFlow       & 90.51                             & 85.67                            & 84.65                    & 79.67          & 77.39          & 73.22          \\
            +AlignFlow(Mask) & \textbf{90.57}                    & \textbf{85.90}                   & \textbf{85.74}           & \textbf{80.59} & \textbf{77.60} & \textbf{73.47} \\

            \bottomrule[2pt]
        \end{tabular}
    }
\end{table}

\noindent\textbf{Implementation Details.} We use the Stable Diffusion 3.5 Medium as the foundation model and employ DINOv3 ViTH16Plus as the feature extraction model when computing the alignment loss and filtering synthetic masks. The same training strategy is adopted across all datasets: all images are resized to $512 \times 512$, and prompts have a 50\% probability of being set as empty strings. We use the AdamW optimizer with a learning rate of $1 \times 10^{-5}$, a weight decay of $1 \times 10^{-2}$, and a batch size of 4. The training is performed on 8 NVIDIA A100 GPUs. We set the training iterations to 4000 for the first stage, and 500 in the second. For the ablation studies, we set the weight of the alignment loss $w_{align}$ to $1.0$ and the start time step index $t_s^i$ to $18$. During inference, we set the guidance scale to $7.0$ and use the Euler Discrete Scheduler\cite{karras2022elucidating} to perform 28 steps.

\subsection{Results}
\label{sec:result}

\begin{figure}[!t]
    \centering
    \centerline{\scalebox{0.5}{\includegraphics[width=\textwidth]{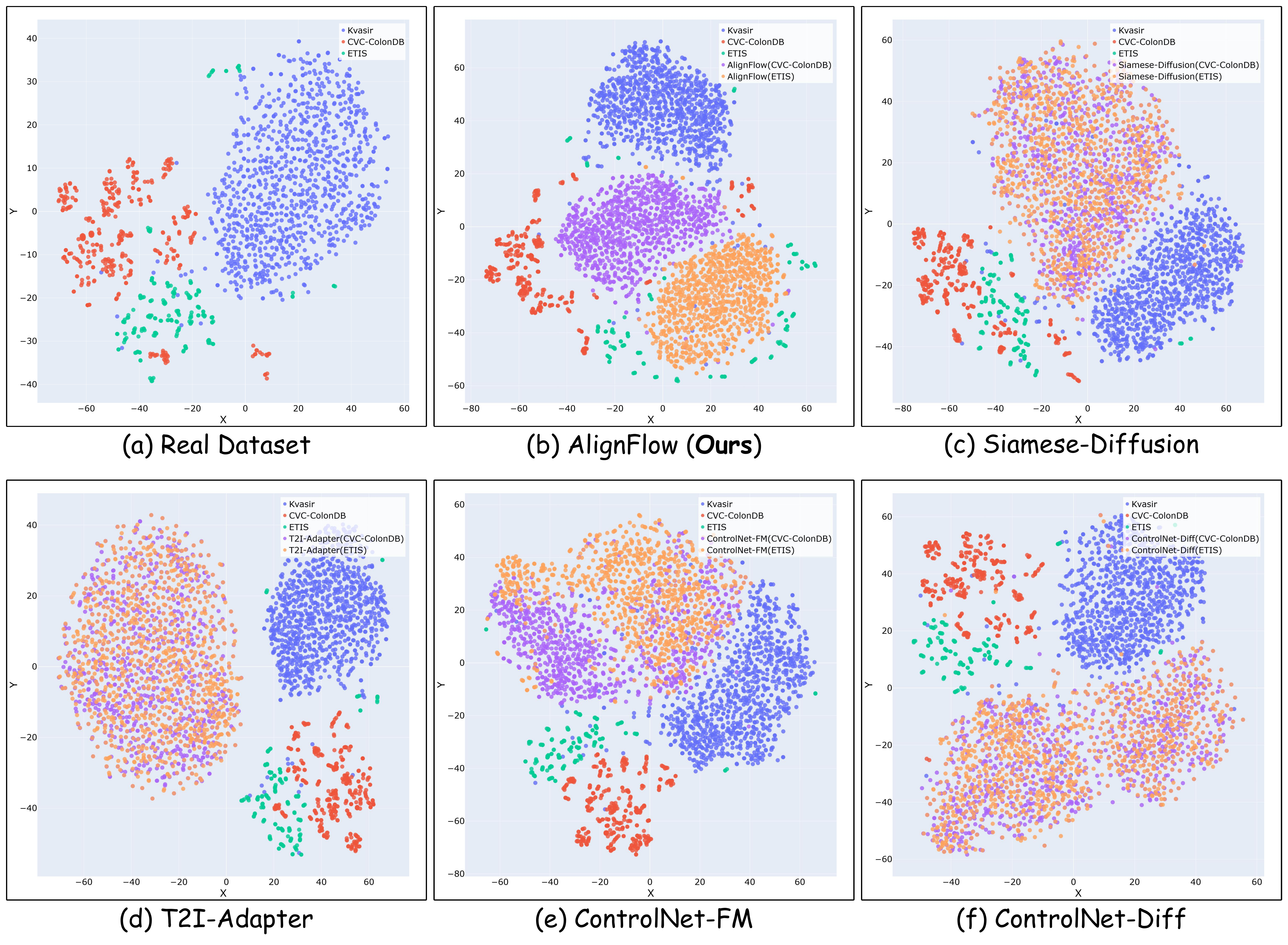}}}
    \caption{t-SNE visualization of the data distributions of images generated by different methods.}
    \label{fig:fig4}
\end{figure}

\noindent\textbf{Quantitative image quality assessment.} In Table \ref{tab:tab1}, we present the experimental results of AlignFlow and other methods on the CVC-ClinicDB dataset. As shown in the table, our method outperforms all other methods across all five evaluation metrics. For the FID and KID metrics, our method achieves 149.27 and 0.1301, which are 7.55 and 0.0043 lower than the second-best Siamese-Diffusion, respectively. This indicates that the images generated by our method are more closely aligned in distribution with the images in the target domain dataset, a conclusion further supported by Figure \ref{fig:fig4}. In addition, our method significantly outperforms others on the traditional LPIPS, SSIM, and PSNR metrics, demonstrating that the images generated by our method are more similar to the target domain dataset in terms of human perception, pixel-level error, brightness, contrast, and structural consistency.

\noindent\textbf{Qualitative image quality assessment.} In Figure \ref{fig:fig3}, we visualize the images generated by AlignFlow and other methods after training on reference images from different target domains. Comparing Figures \ref{fig:fig3}(a) and \ref{fig:fig3}(c), we observe that the images generated by ControlNet-Diff are structurally inconsistent with the images from the source domain. The positions of the optic cup and optic disc are noticeably shifted to the right, indicating poor alignment with the mask. Comparing Figures \ref{fig:fig3}(g) and \ref{fig:fig3}(e), we observe that Siamese-Diffusion produces images with a grayish-white background, which is inconsistent with images from the target domain. Comparing Figure \ref{fig:fig3}(g) with Figures \ref{fig:fig3}(b–f), the images generated by AlignFlow not only align closely with the given mask in structure but also show higher similarity to the target domain images in terms of color, texture, and style, compared to priors.

\noindent\textbf{Visualization of data distribution.} We use the t-SNE\cite{maaten2008visualizing} algorithm to visualize the distributions of images generated by different methods. As shown in Figure \ref{fig:fig4}(b), the images generated by AlignFlow accurately fall within the center of the target domain dataset distribution, indicating that the generated images are consistent with the target domain images in terms of distribution. Comparing Figure \ref{fig:fig4}(b) with Figures \ref{fig:fig4}(c–f), we observe that the images generated by other methods not only exhibit a significant distribution gap from the target domain images, but also show no substantial distributional differences when generated under different reference image guidance. This indicates that other methods fail to utilize the reference images effectively.

\noindent\textbf{Comparison study of model backbones.} We conduct extensive comparative experiments on three commonly used segmentation backbones: UNet\cite{ronneberger2015u}, SegFormer\cite{xie2021segformer}, and DPT\cite{ranftl2021vision}, with the results shown in Table \ref{tab:tab2}. “AlignFlow(SKL)” and “AlignFlow” denote the variants that use $r_{align}^{SKL}$ and $r_{align}^{MMD}$ as the reward function, respectively. “AlignFlow(Mask)” indicates the introduction of new synthetic data on top of the “AlignFlow” setup. Both the masks  and images are generated using our proposed method. The size of the synthetic data is equal to the training set.

On the SegFormer model, AlignFlow achieves the best results across all datasets. Specifically, on the ETIS dataset, AlignFlow attains mDice and mIoU scores of 84.57\% and 80.64\%, respectively, which are 1.59\% and 1.61\% higher than the second-best method, Siamese-Diffusion. On the UNet and DPT models, AlignFlow also achieves the best results on most datasets, with only a few datasets showing slightly lower performance compared to other methods (with differences generally below 0.1\%). These results indicate that our method consistently delivers superior segmentation performance in the majority of cases, demonstrating strong generalizability and robustness.

Comparing AlignFlow(SKL) with AlignFlow, AlignFlow achieves higher metrics in nearly all cases, indicating that our reward function $r_{align}^{MMD}$ is more effective than $r_{align}^{SKL}$; thus, we use $r_{align}^{MMD}$ as the default.

In addition, we validate the effectiveness of our proposed mask synthesis method, as shown by “AlignFlow(Mask)” in Table \ref{tab:tab2_1}. Using synthetic mask data improves performance on most datasets compared with not using it, indicating that our diverse synthesized masks further enhance model generalization.

\begin{table}
    \caption{Comparison of the impact of different number of reference images on medical image segmentation performance, evaluated using mDice (\%) and mIoU (\%) metrics. The numbers in parentheses indicate the number of reference images used.}
    \label{tab:tab3}
    \centering
    \setlength{\tabcolsep}{1.0mm}
    \scalebox{0.51}
    {
        \begin{tabular}{l|c c c c c c|c c c c c c}
            \toprule[2pt]
            \multirow{4}{*}{Methods}                           &
            \multicolumn{6}{c|}{FedPolyp\cite{chen2024fedevi}} & \multicolumn{6}{c}{REFUGE2\cite{fang2022refuge2}}                                                                                                                                                                                                                                                                              \\
            \cmidrule{2-13}
            ~                                                  & \multicolumn{2}{c|}{CVC-ClinicDB}                 & \multicolumn{2}{c|}{CVC-ColonDB} & \multicolumn{2}{c|}{ETIS} & \multicolumn{2}{c|}{KOWA} & \multicolumn{2}{c|}{TOPCON} & \multicolumn{2}{c}{Zeiss}                                                                                                                         \\
            \cmidrule{2-13}
            ~                                                  & mDice                                             & mIoU                             & mDice                     & mIoU                      & mDice                       & mIoU                      & mDice             & mIoU              & mDice             & mIoU              & mDice             & mIoU              \\

            \midrule[1pt]
            UNet                                               & 88.33                                             & 83.11                            & 79.07                     & 74.16                     & 75.46                       & 71.32                     & 89.29             & 82.14             & 84.64             & 76.96             & 87.05             & 79.23             \\
            \hline
            +AlignFlow(1)                                      & 88.14                                             & 82.92                            & 79.11                     & 74.10                     & 76.10                       & 72.14                     & 87.87             & 80.22             & 88.85             & 81.78             & 89.65             & 82.61             \\
            +AlignFlow(5)                                      & 88.56                                             & 83.75                            & 81.41                     & 76.24                     & 75.82                       & 72.18                     & 88.72             & 81.36             & 87.60             & 80.89             & 89.44             & 82.52             \\
            +AlignFlow(10)                                     & \underline{88.68}                                 & \underline{83.77}                & \underline{82.41}         & \underline{77.38}         & \underline{76.45}           & \underline{72.83}         & \underline{89.77} & \underline{82.77} & \textbf{89.14}    & \textbf{82.54}    & \underline{89.94} & \textbf{83.22}    \\
            +AlignFlow(all)                                    & \textbf{88.89}                                    & \textbf{84.03}                   & \textbf{82.90}            & \textbf{77.78}            & \textbf{76.97}              & \textbf{73.27}            & \textbf{90.04}    & \textbf{83.20}    & \underline{89.09} & \underline{82.26} & \textbf{89.95}    & \underline{83.19} \\

            \midrule[1pt]
            SegFormer                                          & 92.15                                             & 87.92                            & 83.53                     & 79.11                     & 81.53                       & 77.83                     & 87.22             & 79.37             & 87.31             & 80.26             & 89.41             & 82.85             \\
            \hline
            +AlignFlow(1)                                      & 92.48                                             & 88.43                            & 84.20                     & 79.30                     & \underline{84.89}           & \underline{80.87}         & 88.98             & 81.82             & 88.27             & 80.94             & 89.61             & 82.71             \\
            +AlignFlow(5)                                      & 92.51                                             & 88.45                            & 86.08                     & 81.05                     & 84.55                       & 80.33                     & 89.48             & 83.00             & 87.20             & 80.64             & 89.32             & 83.00             \\
            +AlignFlow(10)                                     & \textbf{92.87}                                    & \textbf{88.89}                   & \underline{86.52}         & \underline{81.34}         & 84.57                       & 80.64                     & \underline{90.36} & \underline{83.96} & \underline{90.17} & \underline{83.68} & \underline{90.19} & \underline{83.85} \\
            +AlignFlow(all)                                    & \underline{92.80}                                 & \underline{88.74}                & \textbf{86.86}            & \textbf{81.91}            & \textbf{85.86}              & \textbf{81.72}            & \textbf{90.51}    & \textbf{84.26}    & \textbf{90.46}    & \textbf{83.79}    & \textbf{90.55}    & \textbf{84.07}    \\

            \midrule[1pt]
            DPT                                                & 90.03                                             & 85.13                            & 81.66                     & 76.61                     & 73.39                       & 69.76                     & 86.48             & 78.28             & 85.29             & 77.46             & 85.59             & 77.54             \\
            \hline
            +AlignFlow(1)                                      & 90.37                                             & 85.54                            & 83.62                     & 78.22                     & \underline{78.56}           & \underline{73.91}         & 83.60             & 74.65             & 83.74             & 75.46             & 85.89             & 78.34             \\
            +AlignFlow(5)                                      & 90.42                                             & 85.59                            & 84.48                     & 79.26                     & 76.76                       & 72.71                     & 88.02             & 80.64             & 85.48             & 77.92             & 85.40             & 77.70             \\
            +AlignFlow(10)                                     & \underline{90.51}                                 & \underline{85.67}                & \underline{84.65}         & \underline{79.67}         & 77.39                       & 73.22                     & \underline{88.59} & \underline{81.25} & \underline{85.70} & \underline{78.23} & \underline{86.73} & \underline{79.09} \\
            +AlignFlow(all)                                    & \textbf{90.67}                                    & \textbf{85.89}                   & \textbf{85.35}            & \textbf{80.08}            & \textbf{78.65}              & \textbf{74.49}            & \textbf{88.98}    & \textbf{81.61}    & \textbf{86.22}    & \textbf{78.80}    & \textbf{87.11}    & \textbf{79.25}    \\

            \bottomrule[2pt]
        \end{tabular}}
\end{table}
\noindent\textbf{Impact of Reference Image Quantity.} Our goal is to generate synthetic images consistent with the target-domain distribution using only a few target-domain reference images, so we evaluate performance under different reference counts. We test AlignFlow with 1, 5, 10, and all available target-domain images as references. The detailed experimental results are in Table \ref{tab:tab3}.

When the number of reference images is not limited, using all available target domain images as references generally achieves the best results. However, in real world scenarios, obtaining many target domain images is often infeasible, so few shot performance is also crucial. From the results of “AlignFlow(5)” to “AlignFlow(10)”, our method shows significant improvements with only 5–10 reference images; with 10 references, performance is already close to “AlignFlow(all)”. For “AlignFlow(1)”, a single reference image can improve segmentation in some cases but may decline in others, since using only one reference introduces substantial uncertainty: its sampling density within the data distribution directly affects the bias of the generated data. Thus, selecting multiple reference images can help suppress this bias.

\begin{table}[!t]
    \centering
    \caption{Comparison of the impact of different values of $t_s^i$ on medical image segmentation performance, evaluated using mDice (\%) and mIoU (\%) metrics.}
    \label{tab:tab4}
    \scalebox{0.73}
    {
        \begin{tabular}{l|c c|c c|c c|c c}
            \toprule[2pt]
            \multirow{4}{*}{$t_s^i$} & \multicolumn{4}{c|}{KOWA} & \multicolumn{4}{c}{Zeiss}                                                                                                                                      \\
            \cmidrule{2-9}
            ~                        & \multicolumn{2}{c|}{UNet} & \multicolumn{2}{c|}{SegFormer} & \multicolumn{2}{c|}{UNet} & \multicolumn{2}{c}{SegFormer}                                                                     \\
            \cmidrule{2-9}
            ~                        & mDice                     & mIoU                           & mDice                     & mIoU                          & mDice          & mIoU           & mDice          & mIoU           \\
            \midrule[0.75pt]
            0                        & 85.52                     & 76.95                          & 86.30                     & 78.10                         & 87.71          & 79.98          & 88.43          & 81.31          \\
            10                       & 89.14                     & 81.96                          & 89.70                     & 83.25                         & 88.62          & 81.31          & 89.29          & 82.69          \\
            15                       & 89.49                     & 82.48                          & 89.24                     & 82.34                         & 89.78          & 82.88          & 90.32          & 84.00          \\
            16                       & 87.40                     & 79.61                          & 89.86                     & 83.20                         & 88.20          & 80.90          & 88.34          & 81.40          \\
            17                       & 89.20                     & 82.08                          & \textbf{90.55}            & \textbf{84.42}                & 87.22          & 79.74          & 89.76          & 83.05          \\
            18                       & \textbf{89.77}            & \textbf{82.77}                 & 90.36                     & 83.96                         & \textbf{89.94} & \textbf{83.22} & 90.19          & 83.85          \\
            19                       & 88.29                     & 80.84                          & 89.62                     & 83.23                         & 88.85          & 81.68          & \textbf{90.73} & \textbf{84.33} \\
            \bottomrule[2pt]
        \end{tabular}}
\end{table}

\begin{table}[!t]
    \centering
    \caption{Comparison of the impact of different values of $w_{align}$ on medical image segmentation performance, evaluated using mDice (\%) and mIoU (\%) metrics.}
    \label{tab:tab5}
    \scalebox{0.69}
    {
        \begin{tabular}{l|c c|c c|c c|c c}
            \toprule[2pt]
            \multirow{4}{*}{$w_{align}$} & \multicolumn{4}{c|}{KOWA} & \multicolumn{4}{c}{Zeiss}                                                                                                                                      \\
            \cmidrule{2-9}
            ~                            & \multicolumn{2}{c|}{UNet} & \multicolumn{2}{c|}{SegFormer} & \multicolumn{2}{c|}{UNet} & \multicolumn{2}{c}{SegFormer}                                                                     \\
            \cmidrule{2-9}
            ~                            & mDice                     & mIoU                           & mDice                     & mIoU                          & mDice          & mIoU           & mDice          & mIoU           \\
            \midrule[0.75pt]
            0.25                         & 89.07                     & 81.87                          & 89.97                     & 83.42                         & 89.03          & 82.06          & 90.05          & 83.35          \\
            0.5                          & 88.62                     & 81.22                          & 89.71                     & 83.07                         & 89.24          & 82.17          & 89.78          & 83.32          \\
            1.0                          & \textbf{89.77}            & \textbf{82.77}                 & \textbf{90.36}            & \textbf{83.96}                & \textbf{89.94} & \textbf{83.22} & \textbf{90.19} & \textbf{83.85} \\
            2.0                          & 88.69                     & 81.37                          & 89.19                     & 82.27                         & 88.72          & 81.70          & 88.81          & 81.89          \\
            4.0                          & 87.85                     & 80.24                          & 89.10                     & 82.30                         & 88.64          & 81.45          & 88.56          & 81.55          \\
            \bottomrule[2pt]
        \end{tabular}}
\end{table}
\noindent\textbf{Ablation Studies.} Tables \ref{tab:tab4} and \ref{tab:tab5} present ablation studies on two key parameters: $t_s^i$ and $w_{align}$. We first set $w_{align} = 1.0$ by default and then evaluated the performance of AlignFlow under different start time step indices $t_s^i$. From Table \ref{tab:tab4}, we observe that, due to our training strategy of randomly sampling the start time step within the range [$t_s^i$, 19], the model demonstrates robust performance across different values of $t_s^i$. Except for a significant performance drop at $t_s^i = 0$, there is little difference in performance for other values. Based on these experimental results, we chose $t_s^i = 18$, as it achieves near-optimal performance in most cases.

In Table \ref{tab:tab5}, we show how the performance of AlignFlow varies with different values of the weight $w_{align}$ in Equation \ref{eq:eq6}. We observe that when $w_{align}$ is set too high, performance declines to some extent. This may be because an excessively large weight on the alignment loss reduces the optimization of the denoising loss, leading to a decrease in the realism of the generated images. Based on the experimental results in Table \ref{tab:tab5}, we set $w_{align} = 1.0$, as it achieves optimal performance in all tested scenarios.

\section{Conclusion}
\label{sec:conclusion}
In this work, we propose a few-shot distribution-aligned flow matching model, AlignFlow, which uses differentiable reward fine-tuning to directly update gradients for optimizing model parameters. It efficiently aligns the distribution of generated images with that of reference images from the target domain and remains effective even when only a small number of reference images are provided. We also design a distribution alignment mechanism and two distinct reward functions for AlignFlow, providing algorithmic support for efficiently and accurately assessing the distribution differences between generated images and reference images. We conduct extensive experiments on six datasets from different domains with three segmentation models, demonstrating the effectiveness of our proposed framework.

\bibliography{example_paper}
\bibliographystyle{icml2026}

\newpage
\appendix
\onecolumn

\section{Preliminary}
Flow matching is a generative method based on Continuous Normalizing Flows, which learns a time-dependent vector field to transform an initial distribution into a target distribution. Given an initial observation $x_0 \sim \pi_0$ and a target observation $x_1 \sim \pi_1$, the transformation process that maps $x_0$ to $x_1$ over time $t \in [0,1]$ can be expressed by the following ordinary differential equation (ODE):
\begin{align}\label{eq:eq1}
    dx_t = v_t(x_t)dt
\end{align}
The value of $x_1$ can be computed by simulating the ODE using the Euler method,
\begin{align}\label{eq:eq2}
    x_{t+\Delta t} = x_t + v_t(x_t) \Delta t, \ (t=0,\Delta t,\dots,1-\Delta t)
\end{align}
To reduce the number of iterative steps required in the denoising process, most methods typically employ a Gaussian conditional optimal transport probability path,
\begin{align}\label{eq:eq3}
    p_t(x_t|x_1) = N(tx_1, (1-t)^2)
\end{align}
Based on the above setup, we have $x_0 \sim N(0,1)$, $x_t = t x_1 + (1-t) x_0$, and $v_t(x_t) = x_1 - x_0$. When training flow matching, a model $v^\theta_t$ is typically used to predict the vector field of the noisy latent at time $t$. The training objective is to minimize the mean squared error between the true vector field $v_t(x_t)$ and the predicted vector field $v^\theta_t$.
\begin{align}\label{eq:eq4}
    L = \mathbb{E}_{t,x_1,x_0}[\Vert v^\theta_t(tx_1 + (1-t)x_0) - (x_1 - x_0) \Vert^2].
\end{align}

\section{Training Algorithm of AlignFlow}
Given the tunable ControlNet $v^\theta$ and the fully frozen parameters of ControlNet $v^{\theta^*}$, VAE encoder $\varepsilon$, VAE decoder $\varepsilon^{-1}$, and DINOv3 $\sigma$, the training algorithm of AlignFlow is shown in Algorithm \ref{alg:alg1}, where $c_p$ represents prompt latents, $I_{in}$ represents training images, $M_{in}$ represents training masks, $I_{ref}$ represents reference images, $t_s$ is the start time step, and $d_t$ is the time step stride.

\begin{algorithm}
    \caption{Training algorithm of AlignFlow.}\label{alg:alg1}
    \begin{algorithmic}[1]
        \FOR {$c_p$, $I_{in}$, $M_{in}$ in training data}
        \STATE Sample: $\epsilon \sim \mathcal{N}(0,I_d)$, $t \sim \mathcal{U}(0,1)$
        \STATE Encode latents: $x_1 = \varepsilon(I_{in})$, $c_m = \varepsilon(M_{in})$
        \STATE Set $x_t = tx_1 + (1-t)\epsilon$
        \STATE Compute $L_D(\theta) = \Vert v^\theta_t(x_t,t,c_m,c_p) - (x_1 - \epsilon) \Vert^2$
        \IF {in Stage 1}
        \STATE $L(\theta) = L_D(\theta)$
        \ELSE
        \STATE Sample: $x_0 \sim \mathcal{N}(0,I_d)$, $s \sim \mathcal{U}(t_s,1)$
        \STATE Set: $x_{t'}=x_0$, $t'=0$
        \FOR {$t'$ in $range(0,s,d_t)$}
        \STATE $v_{t'} = v^{\theta^*}_{t'}(x_{t'},t',c_m,c_p)$
        \STATE $x_{t'} = x_{t'} + d_t \cdot v_{t'}$
        \ENDFOR
        \STATE $v_{t'} = v^{\theta}_{t'}(x_{t'},t',c_m,c_p)$
        \STATE $\hat{x_0} = x_{t'} + (1-s)v_{t'}$
        \STATE Generate images $I_{out} = \varepsilon^{-1}(\hat{x_0})$
        \STATE Extract features:
        \STATE $F_{out} = \sigma(I_{out})$, $F_{ref} = \sigma(I_{ref})$
        \STATE $L_A(\theta) = MMD(F_{out},F_{ref})$
        \STATE $L(\theta) = L_D(\theta) + w_{align}L_A(\theta)$
        \ENDIF
        \STATE Update the model parameters $\theta$ via gradient descent on $L(\theta)$.
        \ENDFOR
    \end{algorithmic}
\end{algorithm}

\begin{table}[b]
    \caption{Image quality evaluation on TOPCON dataset.}
    \label{tab:tab6}
    \centering
    \setlength{\tabcolsep}{1.2mm}
    \scalebox{0.77}
    {
        \begin{tabular}{l|c c c c c}
            \toprule[2pt]
            Method                                & FID($\downarrow$) & KID($\downarrow$) & LPIPS($\downarrow$) & SSIM($\uparrow$) & PSNR($\uparrow$) \\
            \hline
            T2I-Adapter\cite{mou2024t2i}          & 176.46            & 0.1982            & 0.3245              & 0.3933           & 19.07            \\
            ControlNet-Diff\cite{zhang2023adding} & 64.96             & 0.0637            & 0.2005              & 0.5812           & 21.92            \\
            ControlNet-FM                         & 54.64             & 0.0578            & 0.1787              & 0.5530           & 23.25            \\
            Siamese-Diffusion\cite{qiu2025noise}  & 70.73             & 0.0646            & 0.2060              & 0.5154           & 21.26            \\
            \textbf{AlignFlow(Ours)}              & \textbf{40.26}    & \textbf{0.0356}   & \textbf{0.1644}     & \textbf{0.5993}  & \textbf{25.96}   \\
            \bottomrule[2pt]
        \end{tabular}}
\end{table}

\begin{table}
    \caption{Image quality evaluation on Zeiss dataset.}
    \label{tab:tab7}
    \centering
    \setlength{\tabcolsep}{1.2mm}
    \scalebox{0.77}
    {
        \begin{tabular}{l|c c c c c}
            \toprule[2pt]
            Method                                & FID($\downarrow$) & KID($\downarrow$) & LPIPS($\downarrow$) & SSIM($\uparrow$) & PSNR($\uparrow$) \\
            \hline
            T2I-Adapter\cite{mou2024t2i}          & 152.02            & 0.2135            & 0.3299              & 0.3783           & 20.31            \\
            ControlNet-Diff\cite{zhang2023adding} & 64.78             & 0.0763            & 0.1887              & \textbf{0.6441}  & 24.61            \\
            ControlNet-FM                         & 72.52             & 0.0908            & 0.2020              & 0.5152           & 22.14            \\
            Siamese-Diffusion\cite{qiu2025noise}  & 66.67             & 0.0776            & 0.2046              & 0.5741           & 23.08            \\
            \textbf{AlignFlow(Ours)}              & \textbf{44.88}    & \textbf{0.0471}   & \textbf{0.1821}     & 0.5871           & \textbf{27.64}   \\
            \bottomrule[2pt]
        \end{tabular}}
\end{table}

\section{Quantitative Image Quality Assessment}
We present the image quality evaluation results of AlignFlow on the gastrointestinal polyp dataset CVC-ClinicDB in Table \ref{tab:tab1}. To further validate the robustness of AlignFlow on different types of data, we also test its generated image quality on retinal fundus image datasets TOPCON and Zeiss, with results shown in Tables \ref{tab:tab6} and \ref{tab:tab7}, respectively. Except for achieving suboptimal results in the SSIM metric on the Zeiss dataset, our method achieves the best performance in all other cases, which further demonstrates its robustness.

\begin{figure*}
    \centering
    \centerline{\scalebox{1.0}{\includegraphics[width=\textwidth]{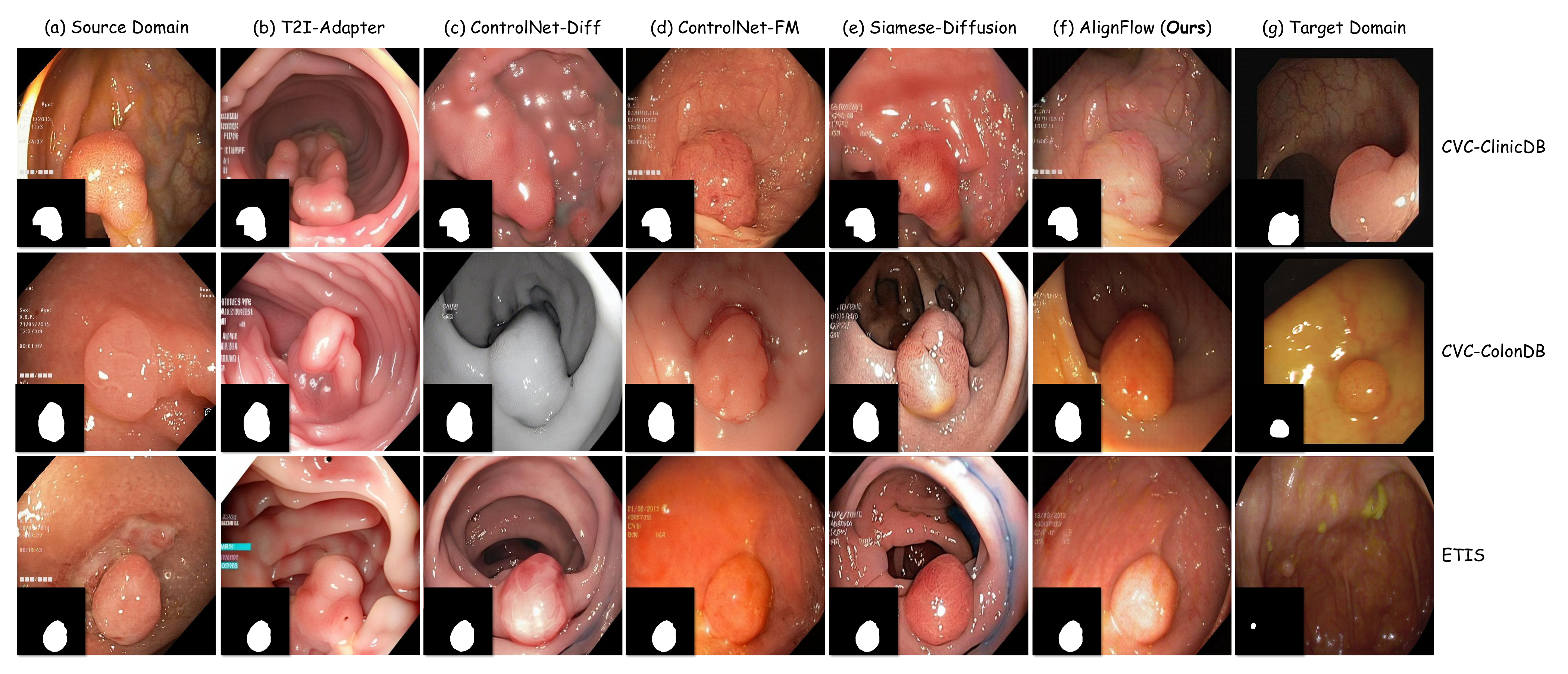}}}
    \caption{Qualitative comparison on FedPolyp dataset. The source domain is Canon, and the target domains are annotated on the right side of each row.}
    \label{fig:fig5}
\end{figure*}

\begin{figure*}
    \centering
    \centerline{\scalebox{1.0}{\includegraphics[width=\textwidth]{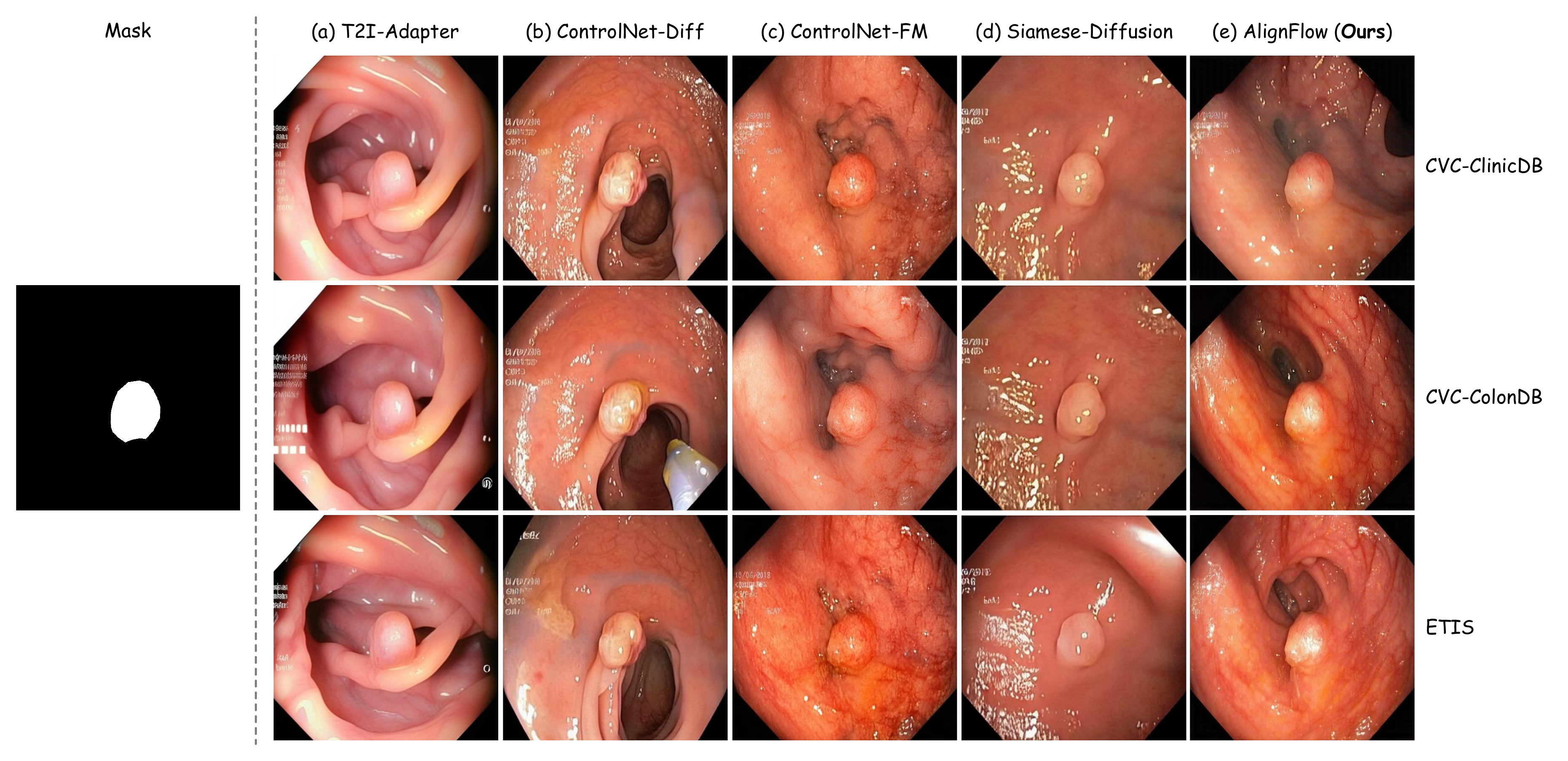}}}
    \caption{Qualitative comparison on FedPolyp dataset using the same mask.}
    \label{fig:fig6}
\end{figure*}

\begin{figure*}
    \centering
    \centerline{\scalebox{1.0}{\includegraphics[width=\textwidth]{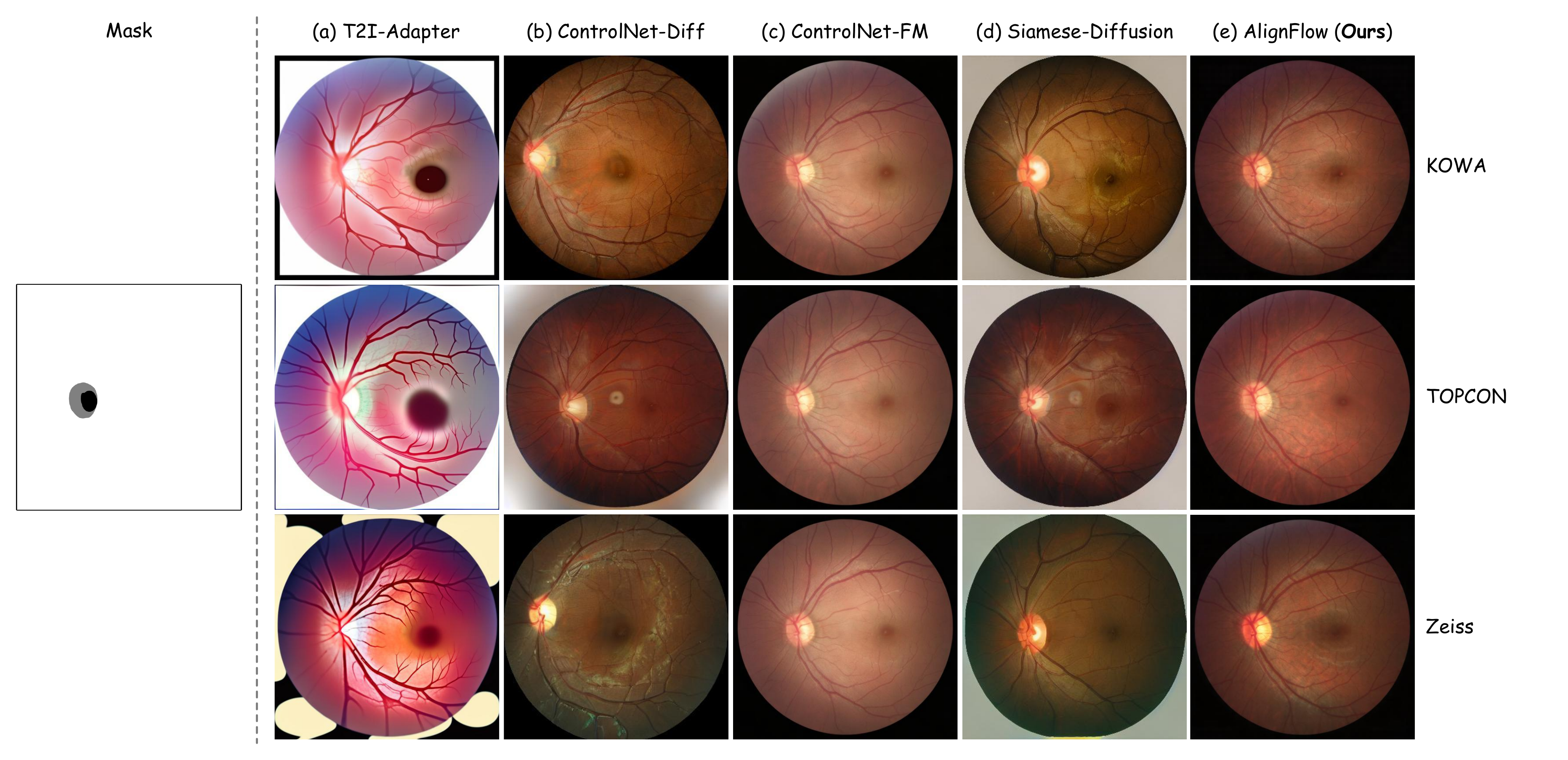}}}
    \caption{Qualitative comparison on REFUGE2 dataset using the same mask.}
    \label{fig:fig7}
\end{figure*}

\section{Qualitative Image Quality Assessment}
As a supplement to Figure \ref{fig:fig3}, we also visualize the images generated by different methods after training on reference datasets from different domains of REFUGE2 in Figure \ref{fig:fig5}. From the figure, we can still observe that the images generated by AlignFlow perform well in terms of structural similarity to the given mask and perceptual similarity to images from the target domain, significantly outperforming other methods. Additionally, we visualize in Figures \ref{fig:fig6} and \ref{fig:fig7} the images generated by different methods after training on reference datasets from different domains, using the same mask for inference. It can be observed that AlignFlow maintains good structural consistency across datasets from different domains.

\end{document}